# The structure of planar defects in tilted perovskites

## RichardBeanland

*Department of Physics, University of Warwick, Coventry, CV4 7AL, UK.*
*E-mail: r.beanland@warwick.ac.uk*

**Synopsis** Continuity of octahedral tilting in perovskite compounds necessarily leads to different tilt structures at planar defects such as domain walls.

**Abstract** A mathematical framework is developed to describe tilted perovskites using a tensor description of octahedral deformations. The continuity of octahedral tilts through the crystal is described using an operator which relates the deformations of adjacent octahedra; examination of the properties of this operator upon application of symmetry elements allows the space group of tilted perovskites to be obtained. It is shown that the condition of octahedral continuity across a planar defect such as an anti-phase boundary or domain wall necessarily leads to different octahedral tilting at the defect, and a method is given to predict the local tilt system which will occur in any given case. Planar boundaries in the rhombohedral $R3c$ $a^-a^-a^-$ tilt system are considered as an example.

**Keywords: Perovskite; octahedral tilting; planar defects; local symmetry**

## 1. INTRODUCTION

Perovskite oxides, of chemical formula $ABO_3$ have been of continuing interest for crystallographers (Lines & Glass, 1979, Glazer, 1972, 1975, O'Keeffe & Hyde, 1977) as well as being technologically important materials, with perovskite compounds exhibiting ferroelectricity, piezoelectricity, giant magneto-resistive and other effects (Lines & Glass, 1979). The prototype perovskite structure is cubic, usually represented as shown in Fig. 1, with A cations at the corners of the unit cell, a B cation at the centre and oxygen anions at the face-centring positions, forming an octahedral cage for the B cation. Typical $ABO_3$ perovskite compounds have the prototype structure at high temperature, while at room temperature the small distortions of the structure and displacements of the atoms within the unit cell are responsible for many of their interesting properties.

In some materials, such as $PbTiO_3$, the distortions which occur upon cooling from the paraelectric cubic phase can be described relatively simply, i.e. by a deformation of the unit cell in combination with displacement of the B cation from its nominal site (Damjanovic, 1998). Others have more complex deformation patterns which are traditionally described by considering 'tilting' of the oxygen octahedra, displacements of the cations, and further distortions of the octahedra.(Glazer, 1972, 1975, O'Keeffe & Hyde, 1977, Megaw & Darlington, 1975, Woodward & Reaney, 2005) Glazer (1972, 1975) gave a description of these crystals using *tilt systems*, describing the rotation of the oxygen octahedra about the [100], [010] and [001] axes. Due to the corner connectivity of the relatively rigid octahedra, the effect of tilting one octahedron about, say, the [100] axis, results in all others in the same (100) plane either tilting in the same, or opposite, direction in a similar manner to a plane of interconnected gears. This results in a doubling of periodicity perpendicular to the tilt axis. Octahedra in a given (100) plane are independent of the (100) planes above and below, and so can either tilt the same sense (in phase) or the opposite sense (anti-phase). In the Glazer notation tilts about the three principal axes are described by a letter, followed by a superscript indicating in-phase (+) or anti-phase (-) tilting. For example, $a^-b^-c^+$ describes a tilt system in which the tilts about all three axes are of different magnitudes, those about [100] and [010] being anti-phase and that about [001] being in-phase. The possible combinations of tilts of different magnitudes about the three different axes and the resulting space group symmetries were described by Glazer (1972, 1975), and considered within the framework of group theory by Aleksandrov (1976) and Howard and Stokes (1998).

Here we present a new mathematical framework which can be used to describe distorted perovskites, and consider in particular the implications of group theoretical analysis for the structure of planar defects in these materials. Rather than consider the distortions to take the form of rotations, we proceed directly to a deformation tensor approach, which has the advantage of mathematical simplicity as well as compatibility with the standard forms of writing symmetry operations. The two approaches – tilts and deformations – give equivalent results, and here we maintain the terminology of octahedral tilting within the framework of our more general approach.

From a group-theoretical analysis of the symmetry of different tilt systems, we proceed to examine the local symmetries which exist at planar defects – particularly anti-phase boundaries and domain walls (twins) – in tilted perovskites. The orientation of low energy twins has been considered by several workers, using lattice matching across interfaces (Zheludev & Shuvalov, 1957, Fousek & Janovec, 1969, Sapriel, 1975), energy mimimisation (Shu & Bhattacharya, 2001) and theories of mechanical twinning (Cahn, 1959). However the effect of octahedral tilts on the local structure of such defects does not seem to have been examined in a systematic way, and is the main purpose of this work. We find that if corner connectivity is maintained, only a few allowed symmetries are possible at the planar defects, which are in general different from the parent space group. We propose that this also gives rise to regions adjacent to the defect which have an intermediate structure, acting as a bridge between the defect structure and the bulk.

## 2. SPACE GROUPS OF TILTED PEROVSKITES

### 2.1. Deformation tensor description of octahedral tilts

An undistorted oxygen octahedron is shown in Fig. 2. We take the origin of the coordinate system to be at its centre; any distortion of the octahedron can be described by considering the vectors $\boldsymbol{\omega}_1$ to $\boldsymbol{\omega}_6$, which give the position of oxygen atoms at the corners of the octahedron in this reference frame. A completely general distortion would require a change in position of all six oxygen atoms, but for simplicity, and to maintain the link to the description in terms of octahedral tilts, we restrict ourselves here to distortions which maintain a centre of symmetry. The octahedron can thus be completely described by the vectors $\boldsymbol{\omega}_1$, $\boldsymbol{\omega}_2$, and $\boldsymbol{\omega}_3$. The distortion of the octahedron can be described using a second rank deformation tensor, i.e.

$$\mathbf{D} = \begin{bmatrix} \frac{\delta\omega_{11}}{\omega_1} & \frac{\delta\omega_{21}}{\omega_2} & \frac{\delta\omega_{31}}{\omega_3} \\ \frac{\delta\omega_{12}}{\omega_1} & \frac{\delta\omega_{22}}{\omega_2} & \frac{\delta\omega_{32}}{\omega_3} \\ \frac{\delta\omega_{13}}{\omega_1} & \frac{\delta\omega_{32}}{\omega_2} & \frac{\delta\omega_{33}}{\omega_3} \end{bmatrix} \qquad (1)$$

and as usual the deformation tensor **D** is linked to the equivalent transformation matrix by **T** = **D** + **I**, where **I** is the identity matrix. Now, a rotation of $\theta$ about the [001] axis is given by

$$\mathbf{T} = \begin{bmatrix} \cos\theta & \sin\theta & 0 \\ -\sin\theta & \cos\theta & 0 \\ 0 & 0 & 1 \end{bmatrix} \quad (2)$$

This can also be written as a deformation tensor using **D** = **T** – **I**, (i.e. the same information in a slightly different form, since subtraction of the identity element does not change the information contained in the matrix). Here, it is useful to split **D** into two components, a symmetric part, describing a strain, and an anti-symmetric part, describing a rotation, i.e.

$$\mathbf{D} = \mathbf{I} - \mathbf{T} = \mathbf{D}_{cell} + \mathbf{D}_{octo} = \begin{bmatrix} 1-\cos\theta & 0 & 0 \\ 0 & 1-\cos\theta & 0 \\ 0 & 0 & 0 \end{bmatrix} + \begin{bmatrix} 0 & c & 0 \\ -c & 0 & 0 \\ 0 & 0 & 0 \end{bmatrix} \quad (3)$$

Where $c = \sin\theta$. In the case of centrosymmetric deformations, all oxygen atoms lie on a plane mid-way between the centres of adjacent octahedra. Thus the symmetric part of **D**, $\mathbf{D}_{cell}$, describes changes in the unit cell dimensions, while the anti-symmetric $\mathbf{D}_{octo}$ describes changes of the oxygen octahedron. The latter part, $\mathbf{D}_{octo}$ is of most importance in determining space group symmetry, and for rotations about all three [100], [010] and [001] axes is

$$\mathbf{D}_{octo} = \begin{bmatrix} 0 & c & -b \\ -c & 0 & a \\ b & -a & 0 \end{bmatrix} \quad (4)$$

If this matrix is derived as a sum of rotations about the three cubic axes, it is necessary to assume that all second order terms are small enough to be ignored, since rotations are in general non-commutative. The typical change in the position of the oxygen atoms in tilted perovskites is roughly 10%, i.e. an oxygen atom with coordinates of [0.5, 0.5, 0] may be displaced to a position [0.45, 0.5, 0]. Second order terms are thus of the order of 1%. However we note that when writing the change in atomic positions as a deformation, it is possible to simply use the changes in atomic coordinates directly without the need for three distinct rotations (which is a very helpful mental construct but has no basis in reality). The oxygen octahedra do not actually rotate; the atoms simply displace from their nominal sites in a coordinated manner. Issues with non-commutivity thus do not arise using the deformation tensor description.

It is clear from Eq. (4) that the three tilts about the three different axes produce independent pure shears *a*, *b* and *c*. It is thus possible to describe the effect of tilts by taking just these three components to form a "tilt vector" **t**, given by

$$\mathbf{t} = \begin{bmatrix} a & b & c \end{bmatrix} \quad (5)$$

Which describes the magnitude and sense of tilts about the [100], [010] and [001] axes for the octahedron at the origin of the coordinate system shown in Fig. 2.

### 2.2. The effect of corner connectivity

The oxygen atoms at the corners of the octahedron shown in Fig. 2 are shared with adjacent octahedra. The displacements of these atoms are thus also part of the deformation tensors describing adjacent octahedra; this is simply a mathematical description of the 'interconnected gear wheel' effect of tilts across a plane perpendicular to the tilt axis. Assuming these deformations to also be centrosymmetric, the deformation of an octahedron at any position **v** can be described by a tilt vector $\mathbf{t}_\mathbf{v}$.

In materials described only by ‘in phase’ or ‘anti-phase’ tilting, only two tilt vectors need to be specified to describe the deformations of all octahedra, corresponding to the octahedron at the origin and a second at [111]. We take the tilt of the octahedron at position **v** = [000] to be given by $\mathbf{t}_{[000]} = [a_1,b_1,c_1]$ and that at **v** = [111] to be given by $\mathbf{t}_{[111]} = [a_2,b_2,c_2]$. It is thus convenient to define a 6-vector $\boldsymbol{\tau} = [\mathbf{t}_{[000]}: \mathbf{t}_{[111]}]$. Tilt systems can be described by the vector $\boldsymbol{\tau}$, e.g. $a^0a^0c^+$ is described by $\boldsymbol{\tau} = [0,0,c: 0,0,c]$, $a^0b^+c^-$ is described by $\boldsymbol{\tau} = [0,b,c: 0,b,-c]$, etc. We note that a different 6-vector representation was used by Howard and Stokes (1998), in which the different elements represented in-phase or anti-phase tilts.

In order to obtain the tilts of an octahedron at a general position $\mathbf{v} = [v_1\ v_2\ v_3]$, we first note that the tilts of two octahedra separated by $[2h\ 2k\ 2l]$, where $h$, $k$, $l$ are integers, are the same. Thus in calculating octahedral tilts a general position vector **v** can be reduced to **v** mod(2), where each $v_i$ is modulo 2, e.g. [4 5 -1] mod(2) = [011]. We now define a compound operator $\mathsf{Q} = (\mathbf{Q}|\mathbf{q})$, composed of a matrix (operator) **Q** and vector **q**, which can be operated upon the location vector {**v** mod(2)} to obtain the tilt vector $\mathbf{t}_v$:

$$\mathbf{Q} = \begin{bmatrix} a_1 + a_2 & 0 & 0 \\ 0 & b_1 + b_2 & 0 \\ 0 & 0 & c_1 + c_2 \end{bmatrix} (-1)^{\mathbf{v}\cdot[111]+1}, \mathbf{q} = \begin{bmatrix} a_1 \\ b_1 \\ c_1 \end{bmatrix} (-1)^{\mathbf{v}\cdot[111]}. \qquad (6)$$

This compound operator uses the familiar Seitz notation (Seitz, 1936) used to describe symmetry operations $\mathsf{S} = (\mathbf{S}|\mathbf{s})$ (Hahn, 2006) and obeys similar multiplication rules, i.e. $\mathsf{QS} = (\mathbf{Q}|\mathbf{q})\bullet(\mathbf{S}|\mathbf{s}) = (\mathbf{QS}|\mathbf{Qs}+\mathbf{q})$. In this notation, the position vector takes the form $\mathsf{V} = (\mathbf{I}|\{\mathbf{v} \bmod(2)\})$, where **I** is the identity operator. Using (6), we find that the tilts of the octahedron $\mathbf{t}_v$ at a general position **v** are given by

$$(\mathbf{Q}|\mathbf{t}_v) = \mathsf{QV} = (\mathbf{Q}|\mathbf{Q}\{\mathbf{v} \bmod(2)\} + \mathbf{q}) \qquad (7)$$

or, more simply, just taking the vector part of (7),

$$\mathbf{t}_v = \mathbf{Q}\{\mathbf{v} \bmod(2)\} + \mathbf{q}. \qquad (8)$$

The operator $\mathsf{Q} = (\mathbf{Q}|\mathbf{q})$ is derived by taking the tilt at **v** = [000] to form **q**, while changing the sign of tilts using $(-1)^{\mathbf{v}.[111]}$ and replacing those tilts appropriately with those at **v** = [111] using **Q**, dependent upon **v**. There are 8 unique tilt vectors in the doubled unit cell given by the vectors **v** mod(2) = [000], [001], [010], [011], [100], [101], [110] and [111]. For example, the octahedron at **v** = [100] has tilts given by $\mathbf{t}_{[100]} = [a_2,-b_1,-c_1]$.

### 2.3. Compatibility of tilts with symmetry operations

The tilts considered in sections 2.1 and 2.2 represent the most general tilt set, in which there is no relation between the magnitude or sign of any of the components $a_1$, $b_1$, $c_1$, $a_2$, $b_2$ and $c_2$. Any symmetry operation which relates the different octahedra will produce constraints on the relative sign and sense of tilts. This can be described by writing eq. (7) in a new reference frame given by the application of the symmetry operator, i.e. replacing $\mathsf{QV}$ with $\mathsf{SQS}^{-1}\mathsf{V}$. Thus the application of a symmetry operator $\mathsf{S} = (\mathbf{S}|\mathbf{s})$ of the prototype cubic spacegroup $\Phi_{cubic}$ changes the tilt vector of the octahedron at position **v** to $\mathbf{t}'_v$, where

$$\mathbf{t}'_v = (\mathbf{SQ}[\{\mathbf{S}^{-1}.(\mathbf{v}-\mathbf{s})\}\mathrm{mod}(2)] + \mathbf{Sq})\ |\mathbf{S}| \qquad (9)$$

where the determinant $|\mathbf{S}|$ has been introduced to take account of the change of sign produced by an improper operation such as a mirror or inversion operator, and the modulo (2) operation is now performed at the displaced origin. In doing this, it is also necessary to modify $\mathsf{Q}$ to take account of the change of origin produced by the translation part **s**, so that $\mathsf{Q}$ is now given by

$$\mathbf{Q} = \begin{bmatrix} a_1 + a_2 & 0 & 0 \\ 0 & b_1 + b_2 & 0 \\ 0 & 0 & c_1 + c_2 \end{bmatrix} (-1)^{(\mathbf{v}+\mathbf{s})\cdot[111]+1}, \quad \mathbf{q} = \begin{bmatrix} a_1 \\ b_1 \\ c_1 \end{bmatrix} (-1)^{(\mathbf{v}+\mathbf{s})\cdot[111]} \qquad (10)$$

This general form of $\mathsf{Q}$ and eq. (9) reduce to that given in eqs. (6) and (8) when $\mathsf{S}$ is the identity operator, i.e. $\mathsf{S} = (\mathbf{I}|\mathbf{0})$. Now, if the operator $\mathsf{S}$ is in the space group of the distorted crystal (i.e. $\mathsf{S} \in \Phi_{crystal}$) then the tilt vector of every octahedron after application of $\mathsf{S}$ must be the same as that in the original crystal, i.e.

$$\mathbf{t}'_v = \mathbf{t}_v,\ \forall\ \mathbf{v}\ \mathrm{mod}(2) \qquad (11)$$

In practice, since the tilts are completely described by **v** = [000] and **v** = [111], the compatibility of a symmetry operation with a given tilt system can be expressed using the 6-vector $\boldsymbol{\tau} = [\mathbf{t}_{[000]}: \mathbf{t}_{[111]}]$, i.e.

$$\text{If} \quad \mathsf{S} \in \Phi_{crystal}, \quad \text{then} \quad \boldsymbol{\tau}' = \boldsymbol{\tau} \qquad (12)$$

For example, the (011) mirror-glide plane given by $\mathsf{S} = (m_{yz}|[100])$ gives

$$\boldsymbol{\tau}' = [-a_2,-c_1,-b_1: -a_1,-c_2,-b_2] = [a_1,b_1,c_1: a_2,b_2,c_2] = \boldsymbol{\tau}, \qquad (13)$$

and this equation is only consistent with tilts which have the form $[a,b_1,-b_1: -a,b_2,-b_2]$, i.e. this symmetry operation can only exist in crystals with tilt systems $a^-b^+b^+$ or $a^-b^-b^-$ (and equivalent tilt systems in which one or more tilts are zero, i.e. $a^0b^+b^+$, $a^0b^-b^-$, $a^-b^0b^0$ and $a^0a^0a^0$).

### 2.4. Space group symmetry of tilted perovskites

The space group of a perovskite with a given tilt system is a subgroup of the prototype space group *Pm*3*m*. The group – sub-group relations between the different tilt systems were given by Howard and Stokes (1998). The number of different *orientational variants* for any given subgroup is given by the ratio of the order of the point groups; for example, the tilt system $a^0b^+b^+$ has point symmetry 4/*mmm* (order 16) and there are three orientational variants which are subgroups of *m*3*m* (order 48). The number of *space group variants* for each tilt system is, in general, smaller than the number of orientational variants (Aizu, 1979) – for example $a^-b^-c^-$ has point symmetry $\bar{1}$ (order 2) with 24 orientational variants; however, all of these orientational variants have the *same* space group $P\bar{1}$, i.e. there is only one space group variant. In the framework described here, it is a simple matter to obtain the elements which are compatible with a given space group variant of a given tilt system by application of equation (12) to all elements in *Pm3m,* selecting only those which are compatible with the 6-vector describing the space group variant of the tilt system. Application of this process to the 23 different tilt systems identified by Glazer (1972) reproduces the 15 space groups obtained by Howard and Stokes (1998). Examples for the symmetry elements ($\mathbf{I}|\mathbf{s}$), ($m_z|\mathbf{s}$) and ($m_{xz}|\mathbf{s}$) are given in Tables 1 – 3, which lists each tilt system, the different space group variants for each, and compatibility with the given symmetry element for all unique values of the translation part **s**. Table 1, with $\mathsf{S}$ = ($\mathbf{I}|\mathbf{s}$), gives the translation group of the crystal, which is of importance in determining the local symmetry at a defect in section 3. We note that the assumption that the deformation of the octahedra is centrosymmetric leads to an equivalence between elements related by the inversion operator, (e.g. mirrors and 2-fold axes) and thus the final space groups are also all centrosymmetric.

## 3. LOCAL SYMMETRY OF PLANAR DEFECTS IN TILTED PEROVSKITES

### 3.1. Definition of a bicrystal and the defect character of an interface

Although the framework outlined in section 2 allows the space group of any given tilt system to be obtained, our main interest here is the nature of 'local' symmetry at any given planar defect in a tilted perovskite oxide. As noted by Pond & Vlachavas (1983), when symmetrically equivalent structures exist which are related by a broken symmetry element $\mathsf{W}$, the defect which lies between them can be characterised by the element $\mathsf{W}$. Here, we call $\mathsf{W}$ the *characteristic symmetry operator* of the defect. This is perhaps most familiar in the case of crystal dislocations, which can be considered to be formed by local breaking of the translation symmetry of the crystal, and are characterised by a translation which is part of the crystal space group, known as the Burgers vector **b**. In tilted perovskite oxides, the different orientational variants (known as domains) are related by the broken symmetry elements which exist in the prototype group *Pm*3*m* but not in the crystal space group. Many of these characteristic symmetry operators describe planar defects (i.e. domain walls) rather than line defects (dislocations).

Here, we formally describe the formation of a planar defect (Pond & Vlachavas, 1983) by defining an interfacial plane with unit normal **n**, passing through the origin of the coordinate system in an infinite crystal which we call the 'white', or $\lambda$ crystal with space group $\Phi_\lambda$. A second, interpenetrating, crystal (which we call the 'black', or $\mu$ crystal) is produced by applying a symmetry operator $\mathsf{W}$ with space group $\Phi_\mu$. Finally atoms in the $\lambda$ crystal are removed from one side of the interfacial plane and atoms in the $\mu$ crystal are removed from the other, giving a bicrystal in which the $\lambda$ and $\mu$ crystals are related by the operator $\mathsf{W}$. Obviously, if $\mathsf{W}$ is an element of the $\lambda$ crystal space group $\Phi_\lambda$, the crystal continuity will be maintained across the interfacial plane and it will have no defect character. Conversely, if $\mathsf{W}$ describes a planar defect, the octahedral tilts on either side of the interfacial plane cannot match *by definition*, since the characteristic symmetry operator $\mathsf{W}$ is not an element of the crystal space groups $\Phi_\lambda$ or $\Phi_\mu$. Therefore, if continuity of oxygen octahedra is maintained, the local symmetry at the defect must have a different structure to that of the bulk. Furthermore, it is also clear that *the local structure at the defect must have a form compatible with the characteristic symmetry operation* $\mathsf{W}$. This is a key observation of this paper, and the consequences of this are examined in more detail below. In what follows, we assume that the deformation of the unit cell, $\mathbf{D}_{cell}$, is dependent upon the octahedral deformation $\mathbf{D}_{octo}$, e.g. if continuity of oxygen octahedra requires all tilts to be zero ($a^0a^0a^0$) the lattice will locally be cubic ($\mathbf{D}_{cell} = \mathbf{0}$). This condition does not have to be satisfied to maintain continuity; but it allows us to maintain a straightforward description of the crystal symmetry.

### 3.2. Compatibility of local and bulk symmetries

When a symmetry operation W = ($\mathbf{W}$|$\mathbf{w}$) is used to describe a planar defect, the operator part $\mathbf{W}$ describes the type of defect (e.g. $\mathbf{W} = m_{yz}$ describes an (011) twin). The component of $\mathbf{w}$ perpendicular to the interface, $(\mathbf{w}\cdot\mathbf{n})\mathbf{n}$, gives the displacement of the interface from the origin at the centre of the oxygen octahedron, while that parallel to the interface describes a rigid-body shift of the $\mu$ crystal with respect to the $\lambda$ crystal. Thus, the operator W = ($m_{yz}$|$\mathbf{0}$) describes an (011) twin which passes through the origin; W = ($m_{yz}$|[011]) describes a twin displaced from the origin, and W = ($m_{yz}$|[100]) describes a combination of a twin and a rigid body shift $\mathbf{w}$ = [100] in the (011) interfacial plane.

Tables 1-3, as well as listing whether the symmetry elements are compatible with any given space group variant, thus also describe planar defects described by translations (i.e. anti-phase boundaries (APBs), Table 1) and twins (on the (011) plane, Table 2; on the (001) plane, Table 3). It can be seen that each symmetry operator W is compatible with a set of tilt systems, Γ(W); for example, the element W = ($m_{yz}$|[100]) is listed in the fourth column of Table 3 and is only found in eight different tilt systems (and variants thereof), i.e.

$$\Gamma(m_{yz}|[100]) = \{a^0a^0a^0,\ a^-b^0b^0,\ a^0b^-b^-,\ a^0b^+b^+,\ a^-a^-a^-,\ a^-b^-b^-,\ a^+a^+a^-,\ a^+a^+c^-\}. \qquad (14)$$

Thus, if oxygen octahedra are continuous across an interface, *no matter what the bulk structure*, the local structure at a domain wall characterised by an (011) twin and a shift of [100] must be one of those given by eq. (14). Nevertheless, in any given bulk structure only a subset of these tilt systems is compatible with a planar defect characterised by W. A translation of the $\lambda$ (or $\mu$) crystal by a vector which is a member of the space group $\Phi_\lambda$ (or $\Phi_\mu$) must, by definition, leave the structure of the interface unchanged. Thus only those tilt systems which are also found in all sets of symmetry operations $\Gamma(\mathbf{W}|\mathbf{w}+\mathbf{u})$, where $\mathbf{u}$ is a lattice translation vector in $\Phi_\lambda$ or $\Phi_\mu$, can characterise planar defects. In other words, the translation group of the local defect symmetry must contain the translation groups of both the $\lambda$ and $\mu$ crystals. The set of tilt systems Λ(W) which are compatible with a given planar defect in a given crystal structure is thus

$$\Lambda(\mathsf{W}) = \Gamma(\mathbf{W}|\mathbf{w}) \cap \Gamma(\mathbf{W}|\mathbf{w} + \mathbf{u}_1) \cap \Gamma(\mathbf{W}|\mathbf{w} + \mathbf{u}_2) \cap ... \qquad (15)$$

Where the $\mathbf{u}_i$ are translation vectors of the $\lambda$ and $\mu$ crystals. In the case of a twin characterised by W = ($m_{yz}$|[100]) in an $a^-a^-a^-$ system, both $\Phi_\lambda$ and $\Phi_\mu$, only contain the translation elements $\mathbf{s}$ = [000], [110], [101] and [011] (Table 1). This leads to the rejection of $a^0b^+b^+$, $a^+a^+a^-$ and $a^+a^+c^-$, since their translation groups do not contain these elements (Table 1), giving the set of tilt systems compatible with W = ($m_{yz}$|[100]) in an $a^-a^-a^-$ system to be

$$\Lambda(m_{yz}|[100]) = \{a^0a^0a^0,\ a^-b^0b^0,\ a^0b^-b^-,\ a^-a^-a^-,\ a^-b^-b^-\} \qquad (16)$$

However this principle can be broken when the interfacial plane lies parallel to {100} and **w** = <100>, normal to the interfacial plane. In this case the interface lies perpendicular to a tilt axis and adjacent layers of octahedra can tilt independently; **w** describes the position of the interfacial plane rather than a rigid body shift of the $\mu$ crystal with respect to the $\lambda$ crystal, and local symmetries can be given by tilt systems in $\Gamma(\mathsf{W})$ rather than $\Lambda(\mathsf{W})$.

Tables 1-3 show that a rich variety of local symmetries is possible for APBs and twins in tilted perovskites, and that for any given case there are often several different tilt systems $\Lambda(\mathsf{W})$ which are compatible with the characteristic symmetry operator $\mathsf{W}$. In any given case, the structure which will form in a real crystal will have the lowest free energy, which cannot be determined by symmetry alone. Furthermore, a gradual transition from one tilt system to the other may be expected across the interface, and in the general case this intermediate structure may have yet another symmetry and energy, different from both the defect itself and the bulk.

Although the structure in any given case must be given by energy considerations, it is a straightforward matter to determine which tilts will change across a planar boundary from equation (9). This can be done by examining the tilt vector of the octahedron at the origin, i.e. at **v** = [000], before ($\lambda$ crystal) and after ($\mu$ crystal) application of the operator $\mathsf{W} = (\mathbf{W}|\mathbf{w})$. In the general case,

$$\mathbf{t}_\lambda = [a_1, b_1, c_1]; \qquad \mathbf{t}_\mu = (\mathbf{WQ[\{W}^{-1}(\mathbf{-w})\}\mathrm{mod}(2)] + \mathbf{Wq})\,|\mathbf{W}|. \qquad (17)$$

For example, in the $R3c$, $a^-a^-a^-$ structure $\boldsymbol{\tau} = [a, a, a\text{: } -a, -a, -a]$; at an anti-phase boundary characterised by $\mathsf{W} = (\mathbf{I}|[100])$, $\mathbf{t}_\lambda = [a, a, a]$ and $\mathbf{t}_\mu = [-a, -a, -a]$, indicating that all three tilts must reverse across the boundary plane.

Taking a simplistic approach, the octahedral tilts at a defect might be described by a tilt vector $\mathbf{t}_{mean}$ which is the mean of that on each side, i.e. $\mathbf{t}_{mean} = (\mathbf{t}_\lambda + \mathbf{t}_\mu)/2$. However it is possible for other local symmetries to exist which are not given by this simple approach, as will be shown in the examples below. Here, rather than give an exhaustive list of defect types and local symmetries in all tilt systems, we consider domain walls in the $a^-a^-a^-$ tilt system as an example. Local symmetries of defects in other tilt systems may be obtained by following a similar procedure.

### 3.3. Example: planar defects in the $a^-a^-a^-$ tilt system

Table 4 lists the five different planar defects that can exist in the $a^-a^-a^-$ tilt system, i.e. APBs, twins on {110} and {100} planes, and combinations of a twin and an APB. This is derived by examination of the appropriate Table for the characteristic symmetry operator, which gives the set of allowable local tilt systems $\Gamma(\mathsf{W})$ as in eq. (14); the smaller set of allowable local tilt systems specific to the $a^-a^-a^-$ tilt system $\Lambda(\mathsf{W})$ is then found by examination of Table 1 and eliminating the tilt systems in $\Gamma(\mathsf{W})$ which do not contain the $\lambda$ and $\mu$ translations. The characteristic symmetry operator is given as $\mathsf{W} = (\mathbf{W}|\mathbf{w} + \mathbf{u})$ to emphasise that the interfacial structure is invariant under application of any translation vector **u** in $\Phi_\lambda$ or $\Phi_\mu$. The tilt vector $\mathbf{t}_\mu$ is given according to eq. (17); a negative value indicates that the sense of tilt reverses across the boundary. The final column gives the local tilt system given by $\mathbf{t}_{mean}$. These different boundaries are shown schematically in Fig. 3. For each, the figure on the left shows a defect in which the bulk structure is maintained up to the interfacial plane, while on the right the network of octahedra is continuous. In each case, the structure which results from continuity of the oxygen octahedra corresponds to $\mathbf{t}_{mean}$. This indicates that the local structure at twins in an $a^-a^-a^-$ structure is likely to have tetragonal ($I4/mcm$) or orthorhombic ($Imma$) symmetry, while that at APBs has the cubic ($Pm\bar{3}m$) form.

The local symmetries which can exist in the special case of an [001] APB or twin which is displaced from the origin (i.e. **w** = [001]) are given in Table 5. The extra degree of freedom afforded by the ability of the oxygen octahedra adjacent to the boundary to rotate independently about [001] leads to the tetragonal $a^0a^0c^+$ structure in the case of APBs and the orthorhombic $a^-a^-c^+$ structure for (001) twins.

## 4. DISCUSSION

The approach outlined in sections 2 and 3 provides a method for predicting the local symmetry at planar defects in tilted perovskites. Since a defect is characterised by a broken symmetry element – which is, by definition, not present in the crystal on either side of the defect – the local structure of the defect must be different to that of the bulk material if octahedral continuity is maintained. It is also clear that rigidity of the oxygen framework must lead to a transition region between the defect and the bulk material on either side; this could be considered to be an 'intermediate' structure which links the two crystals. We have chosen the $a^-a^-a^-$ bulk structure as an example, but this approach could be applied to any tilted perovskite.

Despite the success of this approach, it is only a partial description of the tilted perovskite structure. Only pure shears of the octahedra have been considered; these are sufficient to reproduce the space group analysis of Glazer (1972) and Howard and Stokes (1998), but in real materials the deformations are often more complex. In particular, loss of centrosymmetry is usual, resulting in many of the more useful properties of these materials, such as ferroelectricity and piezoelectricity. Furthermore, small deformations of the unit cell, consistent with the space group imposed by octahedral deformations, are common. Here we have given a description in which $\mathbf{D}_{cell}$ is assumed to be consistent with the symmetry dictated by octahedral tilting, although this does not have to be the case in general.

Nevertheless, the principle of continuity of the octahedra leads to clear predictions of local structure in section (3) which can be compared with experimental observations where available. Furthermore, it has been known for some time that local symmetries can be present at planar defects which do not exist in bulk material (Salje & Zhang, 2009, Jaffe *et al.*, 1971). Models of local structure have been given for some systems, although these appear to have been arrived at by inspection of the individual cases rather than the use of group theory. The loss of translation symmetry elements given in Table I gives rise to APBs in many tilt systems; experimentally, these defects tend to have a meandering structure indicating little dependence of interfacial energy upon crystallographic orientation.(Glazer, 1972, Cheng *et al.*, 2006, Ding & Liang, 2002, Liang *et al.*, 2003, Chen *et al.*, 2001, Lebedev *et al.*, 1998) This implies that the local structure is usually of the $a^0a^0a^0$ form rather than $a^0a^0c^+$, since the latter would lead to strong {001} faceting of APBs. Chen et al. (2001) discussed the effect of APBs on colossal magnetoresistance in $La_{1-x}Ca_xMnO_3$, giving a model of distorted octahedra at the defect in this $a^+b^-b^-$ structure. Similarly, Ding and Liang (2002) and Liang *et al.*, (2003) noted that local distortions are required for continuity of oxygen octahedra at APBs in layered perovskites and $La_{2/3}Ca_{1/3}MnO_3$ respectively. In the latter study, high resolution transmission electron microscope (HRTEM) images showed APBs to have a finite width, and their images are consistent with an $a^0a^0a^0$ local structure. In the $Na_{0.5}Bi_{0.5}TiO_3$ $a^-a^-a^-$ system, Dorcet and co-workers (Dorcet & Trolliard, 2008, Dorcet *et al.*, 2009, Dorcet *et al.*, 2008) proposed a model of (100) twins which corresponds to the W = ($m_z$|[001]) defect in the system used here (Fig. 3, Table 5).

It has also been noted that interactions between different planar defects occur. For example, Ricote et al (2000) noted reduced mobility of {011} domain walls in *R3c* $a^-a^-a^-$ $PbZr_xTi_{1-x}O_3$, and similarly Eitel and Randall (2007) noted interactions between APBs and {011} twins in $PbZr_{0.3}Ti_{0.7}O_3$, leading to pinning of domain walls and changes in macroscopic parameters such as the Rayleigh slope parameter as the material transformed from untilted *R3m* to the *R3c* $a^-a^-a^-$ structure. Here, this can be understood by the different structures of the (011) twin (with $a^0b^-b^-$ tilts) and the (011) twin + APB (with $a^-a^0a^0$ tilts**)**.

Finally, the additional condition of octahedral continuity may have implications for the understanding of morphotropic phase boundaries (Viehland, 2000a, Jin *et al.*, 2003a, b, Ahart *et al.*, 2008), where domain sizes can shrink to sizes of a few nanometres or less (Schmitt *et al.*, 2007, Schönau *et al.*, 2007). The different structures which must exist at defects, as well as the intermediate structure between them and bulk material, may be apparent as a change of global symmetry of the material in a similar manner, or in addition to, adaptive phases (Viehland, 2000b, Jin *et al.*, 2003b, a).

## 5. CONCLUSION

We have presented a new mathematical framework which can be used to describe the structure of tilted perovskites, using a deformation tensor description of the oxygen octahedra and an operator Q which describes connectivity between adjacent octahedra. This allows the space group symmetry of tilted perovskites to be derived and reproduces that of Howard and Stokes (1998). Here, we have used this framework to examine the structure of planar defects in tilted perovskites. We find that the condition of continuity of oxygen octahedra across a planar defect leads to restrictions on the local structure at the boundary, giving a set of allowed tilt systems which are *necessarily* different from the bulk structure. We have considered the rhombohedral *R3c* $a^-a^-a^-$ structure as an example, and find that APBs on general planes have $a^0a^0a^0$ tilts while those on particular (001) planes can have $a^-a^-c^+$ tilts. A twin (domain wall) lying on an (011) plane has $a^0b^-b^-$ tilts, while a combination of an (011) twin + APB has $a^-a^0a^0$ tilts. Twins on (001) planes have $a^0a^0c^-$ or $a^-a^-c^+$ tilts, and an (001) twin + APB has $a^-a^-a^0$ tilts. The implications for interactions between different types of planar defect and domain wall pinning have been briefly discussed.

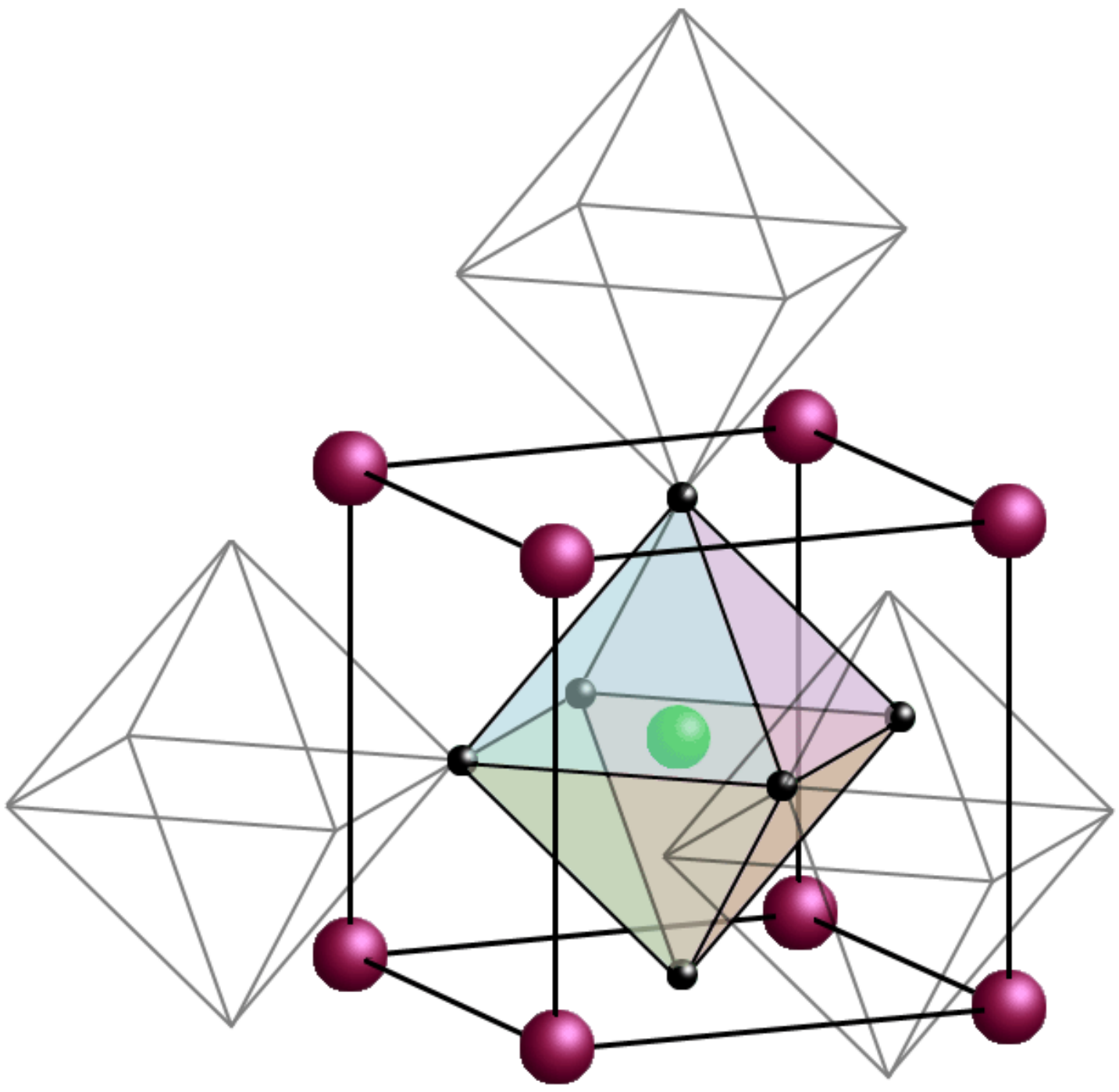

**Figure 1** Schematic of the perovskite structure, showing the A cations (purple), B cations (green) and oxygen (black). The oxygen octahedra, centred on the B cation, is highlighted as well as three adjacent octahedra which are connected at their corners.

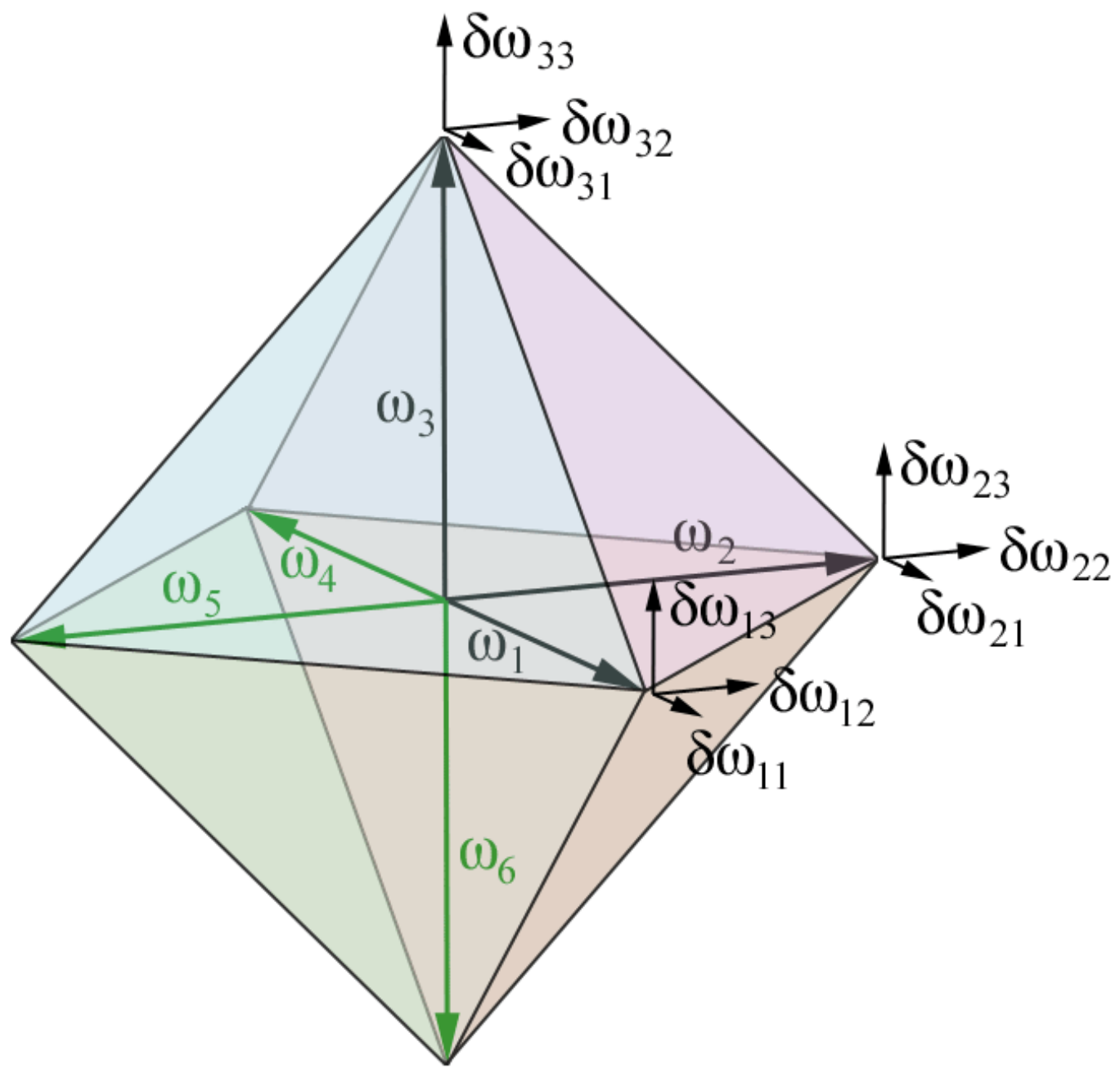

**Figure 2** Deformation of a regular oxygen octahedron, described by three vectors $\boldsymbol{\omega}_i$ and the deformations $\delta\boldsymbol{\omega}_{ij}$.

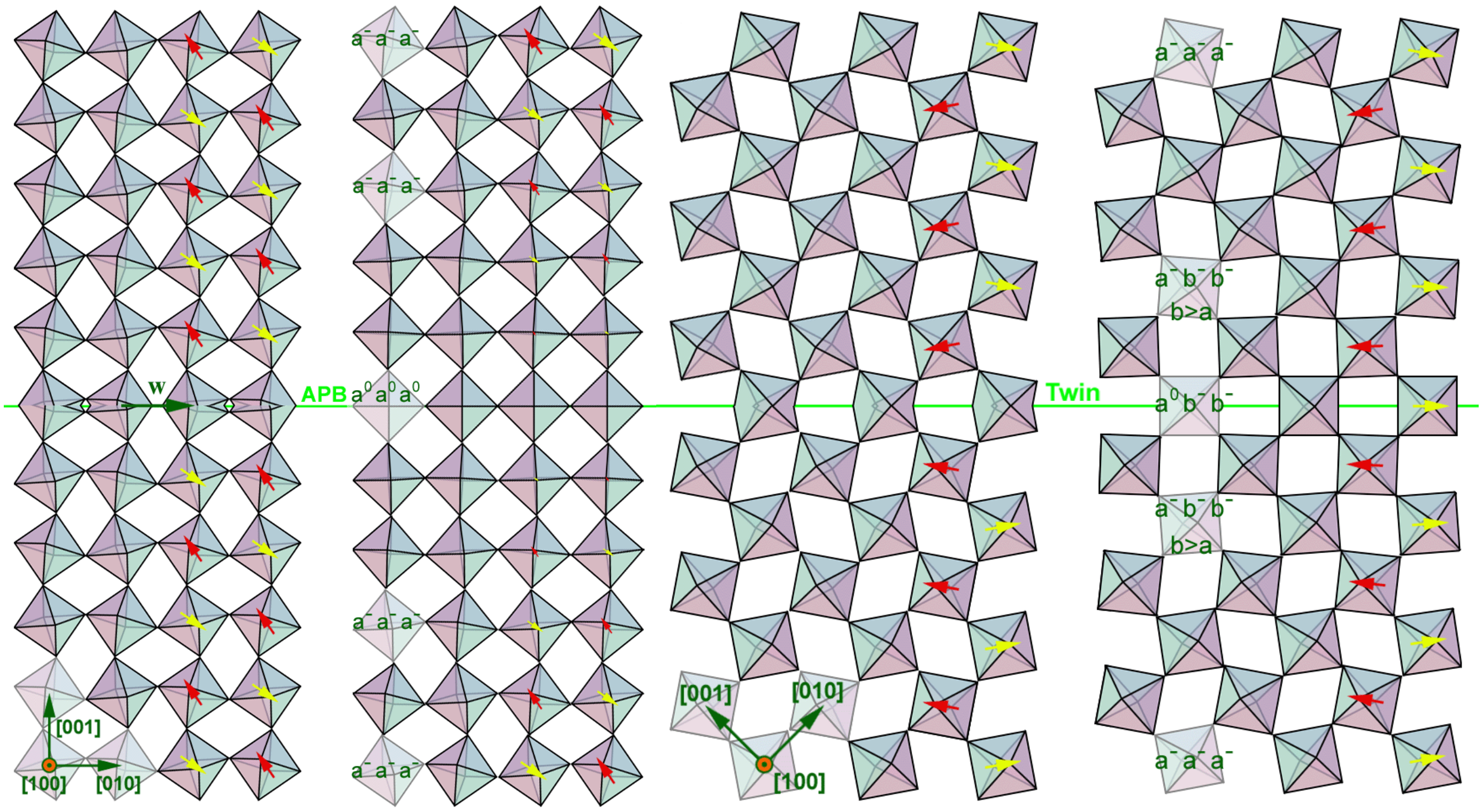

**Figure 3** Schematic showing local octahedral tilting at planar defects in bulk $a^-a^-a^-$ material. Only one (100) plane of oxygen octahedra is shown for clarity, and arrows indicate the tilts about the [010] and [001] axes. (a) an anti-phase boundary (APB) on a (001) plane, W = ($\mathbf{I}$|[010]); (b) an (011) twin, W = ($m_{yz}$|[000]); (c) an (001) twin, W = ($m_z$|[000]), (d) an (011) twin+ APB, W = ($m_{yz}$|[010]) and (e) an (001) twin+ APB, W = ($m_z$|[010]). For each, the structure on the left maintains the $a^-a^-a^-$ structure up to the interface, while that on the right maintains continuity of the oxygen octahedra. The local symmetry at the defects corresponds to that shown in Table 4. Indicated axes show the prototype unit cell; the octahedron at the origin has tilts $\mathbf{t}_\lambda = [a, a, a]$.

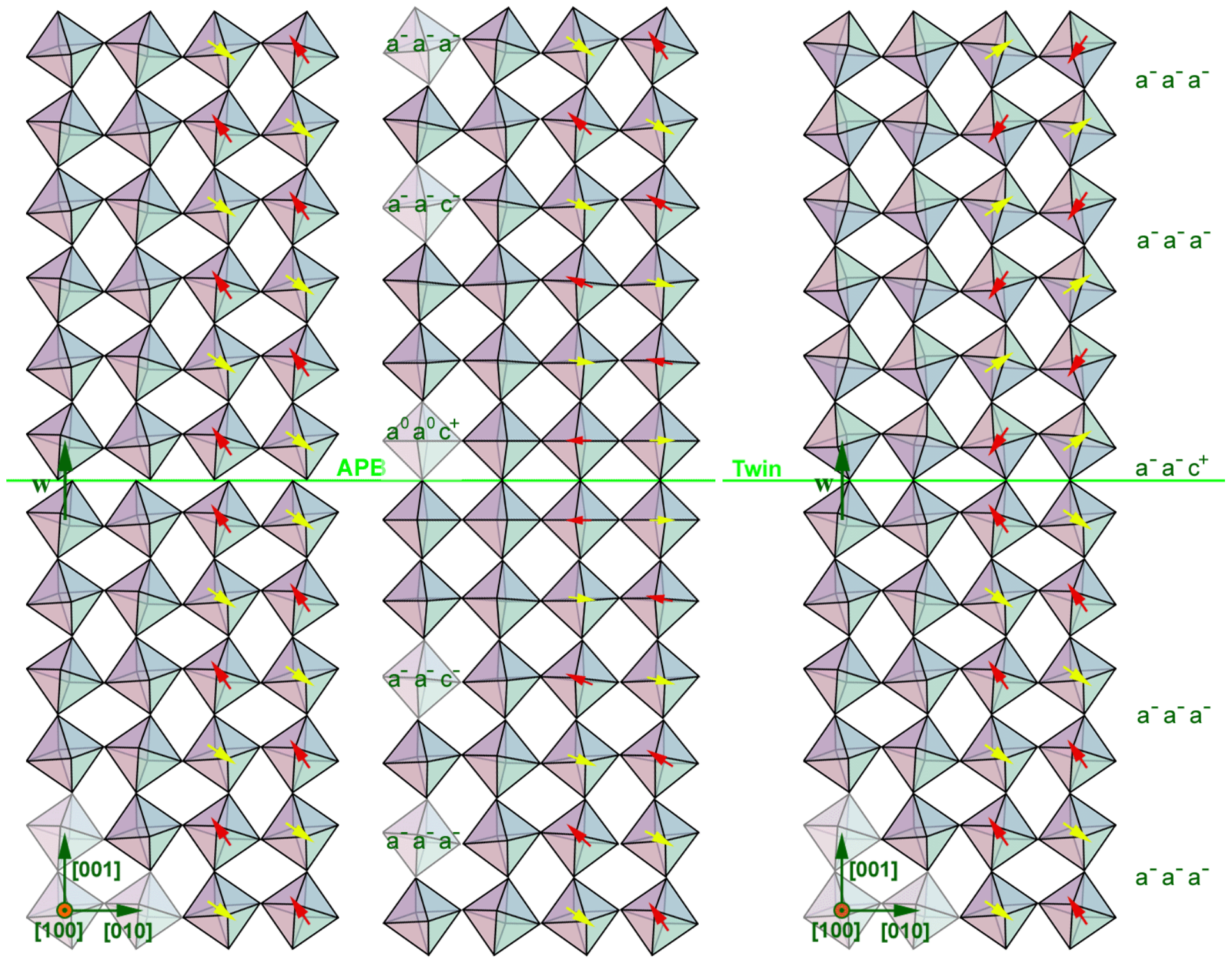

**Figure 4** Schematic showing local octahedral tilting at (001) planar defects in a-a-a- material which pass through the corners of the octahedra. Only one (100) plane of oxygen octahedra is shown for clarity. (a) An APB, W = (I|[001]) (b) a twin W = (mz|[001]). The structure on the left maintains the a-a-a- structure up to the interface. The structure on the right maintains continuity of the oxygen octahedra; the local symmetry is (a) a0a0c+ and (b) a-a-c+. Arrows indicate the tilts about the [010] and [001] axes.

**Table 1** Compatibility of the translation symmetry elements S = (**I**|**s**) with the 23 different tilt systems given by Glazer (1972) and their space group variants. The first column gives the space group and number given by Howard and Stokes (1998). An asterisk indicates that the tilt system has a higher symmetry than that required by the space group, and is therefore not in their main list of isotropy subgroups of Pm3m. The second column gives the number and tilt system from Glazer (1972). The third column gives the 6-vector **τ** = [$\mathbf{t}_{[000]}$: $\mathbf{t}_{[111]}$] used in this work.

**Table 1** Symmetry element S = (**I** | **s**)

| Howard & Stokes | | Glazer | | 6-vector | s | | | | | | | |
|---|---|---|---|---|---|---|---|---|---|---|---|---|
| | | | | | [000] | [001] | [010] | [100] | [011] | [101] | [110] | [111] |
| 1 | $Pm\bar{3}m$ (#221) | 23 | $a^0a^0a^0$ | 0,0,0: 0,0,0 | ✓ | ✓ | ✓ | ✓ | ✓ | ✓ | ✓ | ✓ |
| 6 | *I4/mcm* (#140) | 22 | $a^0a^0c^-$ | 0,0,*c*: 0,0,-*c* | ✓ | | | | ✓ | ✓ | ✓ | |
| | | | | 0,*b*,0: 0,-*b*,0 | ✓ | | | | ✓ | ✓ | ✓ | |
| | | | | *a*,0,0: -*a*,0,0 | ✓ | | | | ✓ | ✓ | ✓ | |
| 2 | *P4/mbm* (#127) | 21 | $a^0a^0c^+$ | 0,0,*c*: 0,0,*c* | ✓ | ✓ | | | | | ✓ | ✓ |
| | | | | 0,*b*,0: 0,*b*,0 | ✓ | | ✓ | | | ✓ | | ✓ |
| | | | | *a*,0,0: *a*,0,0 | ✓ | | | ✓ | ✓ | | | ✓ |
| 7 | *Imma* (#74) | 20 | $a^0b^-b^-$ | 0,*b*,*b*: 0,-*b*,-*b* | ✓ | | | | ✓ | ✓ | ✓ | |
| | | | | 0,*b*,-*b*: 0,-*b*,*b* | ✓ | | | | ✓ | ✓ | ✓ | |
| | | | | *a*,0,*a*: -*a*,0,-*a* | ✓ | | | | ✓ | ✓ | ✓ | |
| | | | | *a*,0,-*a*: -*a*,0,*a* | ✓ | | | | ✓ | ✓ | ✓ | |
| | | | | *a*,*a*,0: -*a*,-*a*,0 | ✓ | | | | ✓ | ✓ | ✓ | |
| | | | | *a*,-*a*,0: -*a*,*a*,0 | ✓ | | | | ✓ | ✓ | ✓ | |
| 9 | *C2/m* (#12) | 19 | $a^0b^-c^-$ | 0,*b*,*c*: 0,-*b*,-*c* | ✓ | | | | ✓ | ✓ | ✓ | |
| | | | | *a*,0,*c*: -*a*,0,-*c* | ✓ | | | | ✓ | ✓ | ✓ | |
| | | | | *a*,*b*,0: -*a*,-*b*,0 | ✓ | | | | ✓ | ✓ | ✓ | |
| * (12) | * *Cmcm* (#63) | 18 | $a^0b^+b^-$ | 0,*b*,*b*: 0,*b*,-*b* | ✓ | | | | | ✓ | | |
| | | | | 0,*b*,*b*: 0,-*b*,*b* | ✓ | | | | | | ✓ | |
| | | | | *a*,0,*a*: *a*,0,-*a* | ✓ | | | | ✓ | | | |
| | | | | *a*,0,*a*: -*a*,0,*a* | ✓ | | | | | | ✓ | |
| | | | | *a*, *a*,0: *a*,-*a*,0 | ✓ | | | | ✓ | | | |
| | | | | *a*, *a*,0: -*a*,*a*,0 | ✓ | | | | | ✓ | | |
| 12 | *Cmcm* (#63) | 17 | $a^0b^+c^-$ | 0,*b*,*c*: 0,*b*,-*c* | ✓ | | | | | ✓ | | |
| | | | | 0,*b*,*c*: 0,-*b*,*c* | ✓ | | | | | | ✓ | |
| | | | | *a*,0,*c*: *a*,0,-*c* | ✓ | | | | ✓ | | | |
| | | | | *a*,0,*c*: -*a*,0,*c* | ✓ | | | | | | ✓ | |
| | | | | *a*,*b*,0: *a*,-*b*,0 | ✓ | | | | ✓ | | | |
| | | | | *a*,*b*,0: -*a*,*b*,0 | ✓ | | | | | ✓ | | |
| 3 | *I4/mmm* (#139) | 16 | $a^0b^+b^+$ | 0,*b*,*b*: 0,*b*,*b* | ✓ | | | | | | | ✓ |
| | | | | 0,*b*,-*b*:0,*b*,-*b* | ✓ | | | | | | | ✓ |
| | | | | *a*,0,*a*: *a*,0,*a* | ✓ | | | | | | | ✓ |
| | | | | *a*,0,-*a*: *a*,0,-*a* | ✓ | | | | | | | ✓ |
| | | | | *a*,*a*,0: *a*,*a*,0 | ✓ | | | | | | | ✓ |
| | | | | *a*,-*a*,0: *a*,-*a*,0 | ✓ | | | | | | | ✓ |
| *(5) | * *Immm* (#71) | 15 | $a^0b^+c^+$ | 0,*b*,*c*: 0,*b*,*c* | ✓ | | | | | | | ✓ |
| 8 | $R\bar{3}c$ (#167) | 14 | $a^-a^-a^-$ | *a*,*a*,*a*: -*a*,-*a*,-*a* | ✓ | | | | ✓ | ✓ | ✓ | |
| | | | | *a*,*a*,-*a*: -*a*,-*a*,*a* | ✓ | | | | ✓ | ✓ | ✓ | |
| | | | | *a*,-*a*,*a*: -*a*,*a*,-*a* | ✓ | | | | ✓ | ✓ | ✓ | |
| | | | | *a*,-*a*,-*a*: -*a*,*a*,*a* | ✓ | | | | ✓ | ✓ | ✓ | |
| 10 | *C2/c* (#15) | 13 | $a^-b^-b^-$ | *a*,*b*,*b*: -*a*,-*b*,-*b* | ✓ | | | | ✓ | ✓ | ✓ | |
| | | | | *a*,*b*,-*b*: -*a*,-*b*,*b* | ✓ | | | | ✓ | ✓ | ✓ | |
| | | | | *a*,*b*,*a*: -*a*,-*b*,-*a* | ✓ | | | | ✓ | ✓ | ✓ | |
| | | | | *a*,*b*,-*a*: -*a*,-*b*,*a* | ✓ | | | | ✓ | ✓ | ✓ | |
| | | | | *a*,*a*,*c*: -*a*,-*a*,-*c* | ✓ | | | | ✓ | ✓ | ✓ | |
| | | | | *a*,-*a*,*c*: -*a*,*a*,-*c* | ✓ | | | | ✓ | ✓ | ✓ | |
| 11 | $P\bar{1}$ (#2) | 12 | $a^-b^-c^-$ | *a*,*b*,*c*: -*a*,-*b*,-*c* | ✓ | | | | ✓ | ✓ | ✓ | |
| * | * | 11 | $a^+a^-a^-$ | *a*,*a*,*a*: *a*,-*a*,-*a* | ✓ | | | | ✓ | | | |
| (13) | *Pnma* (#62) | | | *a,a,-a*: *a,-a,a* | ✓ | | | | ✓ | | | |
| | | | | *a,a,a*: *-a,a,-a* | ✓ | | | | | ✓ | | |
| | | | | *a,a,-a*: *-a,a,a* | ✓ | | | | | ✓ | | |
| | | | | *a,a,a*: *-a,-a,a* | ✓ | | | | | | ✓ | |
| | | | | *a,-a,a*: *-a,a,a* | ✓ | | | | | | ✓ | |
| 13 | *Pnma* (#62) | 10 | $a^+b^-b^-$ | *a,b,b: a,-b,-b* | ✓ | | | | ✓ | | | |
| | | | | *a,b,-b: a,-b,b* | ✓ | | | | ✓ | | | |
| | | | | *a,b,a: -a,b,-a* | ✓ | | | | | ✓ | | |
| | | | | *a,b,-a: -a,b,a* | ✓ | | | | | ✓ | | |
| | | | | *a,a,c: -a,-a,c* | ✓ | | | | | | ✓ | |
| | | | | *a,-a,c: -a,a,c* | ✓ | | | | | | ✓ | |
| * (14) | * $P2_1/m$ (#11) | 9 | $a^+a^-c^-$ | *a,a,c: a,-a,-c* | ✓ | | | | ✓ | | | |
| | | | | *a,b,b: -a,b,-b* | ✓ | | | | | ✓ | | |
| | | | | *a,b,a: -a,-b,a* | ✓ | | | | | | ✓ | |
| 14 | $P2_1/m$ (#11) | 8 | $a^+b^-c^-$ | *a,b,c: a,-b,-c* | ✓ | | | | ✓ | | | |
| | | | | *a,b,c: -a,b,-c* | ✓ | | | | | ✓ | | |
| | | | | *a,b,c: -a,-b,c* | ✓ | | | | | | ✓ | |
| * (15) | * $P4_2/nmc$ (#137) | 7 | $a^+a^+a^-$ | *a,a,a: a,a,-a* | ✓ | | | | | | | |
| | | | | *a,-a,a: a,-a,-a* | ✓ | | | | | | | |
| | | | | *a,a,a: a,-a,a* | ✓ | | | | | | | |
| | | | | *a,a,-a: a,-a,-a* | ✓ | | | | | | | |
| | | | | *a,a,a: -a,a,a* | ✓ | | | | | | | |
| | | | | *a,a,-a: -a,a,-a* | ✓ | | | | | | | |
| * | * *Pmmn* (#59) | 6 | $a^+b^+b^-$ | *a,b,b: a,b,-b* | ✓ | | | | | | | |
| | | | | *a,b,a: a,b,-a* | ✓ | | | | | | | |
| | | | | *a,a,c: a,-a,c* | ✓ | | | | | | | |
| 15 | $P4_2/nmc$ (#137) | 5 | $a^+a^+c^-$ | *a,a,c: a,a,-c* | ✓ | | | | | | | |
| | | | | *a,-a,c: a,-a,-c* | ✓ | | | | | | | |
| | | | | *a,b,a: a,-b,a* | ✓ | | | | | | | |
| | | | | *a,b,-a: a,-b,-a* | ✓ | | | | | | | |
| | | | | *a,b,b: -a,b,b* | ✓ | | | | | | | |
| | | | | *a,b,-b: -a,b,-b* | ✓ | | | | | | | |
| * | * *Pmmn* (#59) | 4 | $a^+b^+c^-$ | *a,b,c: a,b,-c* | ✓ | | | | | | | |
| | | | | *a,b,c: a,-b,c* | ✓ | | | | | | | |
| | | | | *a,b,c: -a,b,c* | ✓ | | | | | | | |
| 4 | $Im\bar{3}$ (#204) | 3 | $a^+a^+a^+$ | *a,a,a: a,a,a* | ✓ | | | | | | | ✓ |
| | | | | *a,a,-a: a,a,-a* | ✓ | | | | | | | ✓ |
| | | | | *a,-a,a: a,-a,a* | ✓ | | | | | | | ✓ |
| | | | | *a,-a,-a: a,-a,-a* | ✓ | | | | | | | ✓ |
| * | * *Immm* (#71) | 2 | $a^+b^+b^+$ | *a,b,b: a,b,b* | ✓ | | | | | | | ✓ |
| | | | | *a,b,a: a,b,a* | ✓ | | | | | | | ✓ |
| | | | | *a,a,c: a,a,c* | ✓ | | | | | | | ✓ |
| 5 | *Immm* (#71) | 1 | $a^+b^+c^+$ | *a,b,c: a,b,c* | ✓ | | | | | | | ✓ |

**Table 2** Compatibility of the mirror symmetry elements S = ($m_{yz}$|**s**) with the 23 different tilt systems given by Glazer (1972) and their space group variants.

**Table 2** Symmetry element S = ($m_{yz}$ | **s**)

| Howard & Stokes | | Glazer | | 6-vector | s | | | | | | | |
|---|---|---|---|---|---|---|---|---|---|---|---|---|
| | | | | | [000] | [001] | [010] | [100] | [011] | [101] | [110] | [111] |
| 1 | $Pm\bar{3}m$ (#221) | 23 | $a^0a^0a^0$ | 0,0,0: 0,0,0 | ✓ | ✓ | ✓ | ✓ | ✓ | ✓ | ✓ | ✓ |
| 6 | *I4/mcm* (#140) | 22 | $a^0a^0c^-$ | 0,0,*c*: 0,0,-*c* | | | | | | | | |
| | | | | 0,*b*,0: 0,-*b*,0 | | | | | | | | |
| | | | | *a*,0,0: -*a*,0,0 | | ✓ | ✓ | ✓ | | | | ✓ |
| 2 | *P4/mbm* (#127) | 21 | $a^0a^0c^+$ | 0,0,*c*: 0,0,*c* | | | | | | | | |
| | | | | 0,*b*,0: 0,*b*,0 | | | | | | | | |
| | | | | *a*,0,0: *a*,0,0 | | ✓ | ✓ | | | ✓ | ✓ | |
| 7 | *Imma* (#74) | 20 | $a^0b^-b^-$ | 0,*b*,*b*: 0,-*b*,-*b* | ✓ | | | | ✓ | ✓ | ✓ | |
| | | | | 0,*b*,-*b*: 0,-*b*,*b* | | ✓ | ✓ | ✓ | | | | ✓ |
| | | | | *a*,0,*a*: -*a*,0,-*a* | | | | | | | | |
| | | | | *a*,0,-*a*: -*a*,0,*a* | | | | | | | | |
| | | | | *a*,*a*,0: -*a*,-*a*,0 | | | | | | | | |
| | | | | *a*,-*a*,0: -*a*,*a*,0 | | | | | | | | |
| 9 | *C2/m* (#12) | 19 | $a^0b^-c^-$ | 0,*b*,*c*: 0,-*b*,-*c* | | | | | | | | |
| | | | | *a*,0,*c*: -*a*,0,-*c* | | | | | | | | |
| | | | | *a*,*b*,0: -*a*,-*b*,0 | | | | | | | | |
| * (12) | * *Cmcm* (#63) | 18 | $a^0b^+b^-$ | 0,*b*,*b*: 0,*b*,-*b* | | | | | | | | |
| | | | | 0,*b*,*b*: 0,-*b*,*b* | | | | | | | | |
| | | | | *a*,0,*a*: *a*,0,-*a* | | | | | | | | |
| | | | | *a*,0,*a*: -*a*,0,*a* | | | | | | | | |
| | | | | *a*, *a*,0: *a*,-*a*,0 | | | | | | | | |
| | | | | *a*, *a*,0: -*a*,*a*,0 | | | | | | | | |
| 12 | *Cmcm* (#63) | 17 | $a^0b^+c^-$ | 0,*b*,*c*: 0,*b*,-*c* | | | | | | | | |
| | | | | 0,*b*,*c*: 0,-*b*,*c* | | | | | | | | |
| | | | | *a*,0,*c*: *a*,0,-*c* | | | | | | | | |
| | | | | *a*,0,*c*: -*a*,0,*c* | | | | | | | | |
| | | | | *a*,*b*,0: *a*,-*b*,0 | | | | | | | | |
| | | | | *a*,*b*,0: -*a*,*b*,0 | | | | | | | | |
| 3 | *I4/mmm* (#139) | 16 | $a^0b^+b^+$ | 0,*b*,*b*: 0,*b*,*b* | ✓ | | | | | | | ✓ |
| | | | | 0,*b*,-*b*:0,*b*,-*b* | | | | ✓ | ✓ | | | |
| | | | | *a*,0,*a*: *a*,0,*a* | | | | | | | | |
| | | | | *a*,0,-*a*: *a*,0,-*a* | | | | | | | | |
| | | | | *a*,*a*,0: *a*,*a*,0 | | | | | | | | |
| | | | | *a*,-*a*,0: *a*,-*a*,0 | | | | | | | | |
| *(5) | * *Immm* (#71) | 15 | $a^0b^+c^+$ | 0,*b*,*c*: 0,*b*,*c* | | | | | | | | |
| 8 | $R\bar{3}c$ (#167) | 14 | $a^-a^-a^-$ | *a*,*a*,*a*: -*a*,-*a*,-*a* | | | | | | | | |
| | | | | *a*,*a*,-*a*: -*a*,-*a*,*a* | | ✓ | ✓ | ✓ | | | | ✓ |
| | | | | *a*,-*a*,*a*: -*a*,*a*,-*a* | | ✓ | ✓ | ✓ | | | | ✓ |
| | | | | *a*,-*a*,-*a*: -*a*,*a*,*a* | | | | | | | | |
| 10 | *C2/c* (#15) | 13 | $a^-b^-b^-$ | *a*,*b*,*b*: -*a*,-*b*,-*b* | | | | | | | | |
| | | | | *a*,*b*,-*b*: -*a*,-*b*,*b* | | ✓ | ✓ | ✓ | | | | ✓ |
| | | | | *a*,*b*,*a*: -*a*,-*b*,-*a* | | | | | | | | |
| | | | | *a*,*b*,-*a*: -*a*,-*b*,*a* | | | | | | | | |
| | | | | *a*,*a*,*c*: -*a*,-*a*,-*c* | | | | | | | | |
| | | | | *a*,-*a*,*c*: -*a*,*a*,-*c* | | | | | | | | |
| 11 | $P\bar{1}$ (#2) | 12 | $a^-b^-c^-$ | *a*,*b*,*c*: -*a*,-*b*,-*c* | | | | | | | | |
| * (13) | * *Pnma* (#62) | 11 | $a^+a^-a^-$ | *a*,*a*,*a*: *a*,-*a*,-*a* | | | | | | ✓ | ✓ | |
| | | | | *a*,*a*,-*a*: *a*,-*a*,*a* | | ✓ | ✓ | | | | | |
| | | | | *a*,*a*,*a*: -*a*,*a*,-*a* | | | | | | | | |
| | | | | *a*,*a*,-*a*: -*a*,*a*,*a* | | | | | | | | |
| | | | | *a*,*a*,*a*: -*a*,-*a*,*a* | | | | | | | | |
| | | | | *a*,-*a*,*a*: -*a*,*a*,*a* | | | | | | | | |
| 13 | *Pnma* (#62) | 10 | $a^+b^-b^-$ | *a,b,b: a,-b,-b* | | | | | | ✓ | ✓ | |
| | | | | *a,b,-b: a,-b,b* | | ✓ | ✓ | | | | | |
| | | | | *a,b,a: -a,b,-a* | | | | | | | | |
| | | | | *a,b,-a: -a,b,a* | | | | | | | | |
| | | | | *a,a,c: -a,-a,c* | | | | | | | | |
| | | | | *a,-a,c: -a,a,c* | | | | | | | | |
| * (14) | * $P2_1/m$ (#11) | 9 | $a^+a^-c^-$ | *a,a,c: a,-a,-c* | | | | | | | | |
| | | | | *a,b,b: -a,b,-b* | | | | | | | | |
| | | | | *a,b,a: -a,-b,a* | | | | | | | | |
| 14 | $P2_1/m$ (#11) | 8 | $a^+b^-c^-$ | *a,b,c: a,-b,-c* | | | | | | | | |
| | | | | *a,b,c: -a,b,-c* | | | | | | | | |
| | | | | *a,b,c: -a,-b,c* | | | | | | | | |
| * (15) | * $P4_2/nmc$ (#137) | 7 | $a^+a^+a^-$ | *a,a,a: a,a,-a* | | | | | | | | |
| | | | | *a,-a,a: a,-a,-a* | | | | | | | | |
| | | | | *a,a,a: a,-a,a* | | | | | | | | |
| | | | | *a,a,-a: a,-a,-a* | | | | | | | | |
| | | | | *a,a,a: -a,a,a* | | | | | | | | ✓ |
| | | | | *a,a,-a: -a,a,-a* | | | | ✓ | | | | |
| * | * *Pmmn* (#59) | 6 | $a^+b^+b^-$ | *a,b,b: a,b,-b* | | | | | | | | |
| | | | | *a,b,a: a,b,-a* | | | | | | | | |
| | | | | *a,a,c: a,-a,c* | | | | | | | | |
| 15 | $P4_2/nmc$ (#137) | 5 | $a^+a^+c^-$ | *a,a,c: a,a,-c* | | | | | | | | |
| | | | | *a,-a,c: a,-a,-c* | | | | | | | | |
| | | | | *a,b,a: a,-b,a* | | | | | | | | |
| | | | | *a,b,-a: a,-b,-a* | | | | | | | | |
| | | | | *a,b,b: -a,b,b* | | | | | | | | ✓ |
| | | | | *a,b,-b: -a,b,-b* | | | | ✓ | | | | |
| * | * *Pmmn* (#59) | 4 | $a^+b^+c^-$ | *a,b,c: a,b,-c* | | | | | | | | |
| | | | | *a,b,c: a,-b,c* | | | | | | | | |
| | | | | *a,b,c: -a,b,c* | | | | | | | | |
| 4 | $Im\bar{3}$ (#204) | 3 | $a^+a^+a^+$ | *a,a,a: a,a,a* | | | | | | | | |
| | | | | *a,a,-a: a,a,-a* | | | | | | | | |
| | | | | *a,-a,a: a,-a,a* | | | | | | | | |
| | | | | *a,-a,-a: a,-a,-a* | | | | | | | | |
| * | * *Immm* (#71) | 2 | $a^+b^+b^+$ | *a,b,b: a,b,b* | | | | | | | | |
| | | | | *a,b,a: a,b,a* | | | | | | | | |
| | | | | *a,a,c: a,a,c* | | | | | | | | |
| 5 | *Immm* (#71) | 1 | $a^+b^+c^+$ | *a,b,c: a,b,c* | | | | | | | | |

**Table 3** Compatibility of the mirror symmetry elements S = ($m_z$|**s**) with the 23 different tilt systems given by Glazer (1972) and their space group variants.

**Table 3** Symmetry element S = ($m_z$ | **s**)

| Howard & Stokes | | Glazer | | 6-vector | s | | | | | | | |
|---|---|---|---|---|---|---|---|---|---|---|---|---|
| | | | | | [000] | [001] | [010] | [100] | [011] | [101] | [110] | [111] |
| 1 | $Pm\bar{3}m$ (#221) | 23 | $a^0a^0a^0$ | 0,0,0: 0,0,0 | ✓ | ✓ | ✓ | ✓ | ✓ | ✓ | ✓ | ✓ |
| 6 | *I4/mcm* (#140) | 22 | $a^0a^0c^-$ | 0,0,*c*: 0,0,-*c* | ✓ | | | | ✓ | ✓ | ✓ | |
| | | | | 0,*b*,0: 0,-*b*,0 | | ✓ | ✓ | ✓ | | | | ✓ |
| | | | | *a*,0,0: -*a*,0,0 | | ✓ | ✓ | ✓ | | | | ✓ |
| 2 | *P4/mbm* (#127) | 21 | $a^0a^0c^+$ | 0,0,*c*: 0,0,*c* | ✓ | ✓ | | | | | ✓ | ✓ |
| | | | | 0,*b*,0: 0,*b*,0 | | ✓ | | ✓ | ✓ | | ✓ | |
| | | | | *a*,0,0: *a*,0,0 | | ✓ | ✓ | | | ✓ | ✓ | |
| 7 | *Imma* (#74) | 20 | $a^0b^-b^-$ | 0,*b*,*b*: 0,-*b*,-*b* | | | | | | | | |
| | | | | 0,*b*,-*b*: 0,-*b*,*b* | | | | | | | | |
| | | | | *a*,0,*a*: -*a*,0,-*a* | | | | | | | | |
| | | | | *a*,0,-*a*: -*a*,0,*a* | | | | | | | | |
| | | | | *a*,*a*,0: -*a*,-*a*,0 | | ✓ | ✓ | ✓ | | | | ✓ |
| | | | | *a*,-*a*,0: -*a*,*a*,0 | | ✓ | ✓ | ✓ | | | | ✓ |
| 9 | *C2/m* (#12) | 19 | $a^0b^-c^-$ | 0,*b*,*c*: 0,-*b*,-*c* | | | | | | | | |
| | | | | *a*,0,*c*: -*a*,0,-*c* | | | | | | | | |
| | | | | *a*,*b*,0: -*a*,-*b*,0 | | ✓ | ✓ | ✓ | | | | ✓ |
| * (12) | * *Cmcm* (#63) | 18 | $a^0b^+b^-$ | 0,*b*,*b*: 0,*b*,-*b* | | | | | ✓ | | ✓ | |
| | | | | 0,*b*,*b*: 0,-*b*,*b* | | ✓ | | | | | | ✓ |
| | | | | *a*,0,*a*: *a*,0,-*a* | | | | | | ✓ | ✓ | |
| | | | | *a*,0,*a*: -*a*,0,*a* | | ✓ | | | | | | ✓ |
| | | | | *a*, *a*,0: *a*,-*a*,0 | | ✓ | ✓ | | | | | |
| | | | | *a*, *a*,0: -*a*,*a*,0 | | ✓ | | ✓ | | | | |
| 12 | *Cmcm* (#63) | 17 | $a^0b^+c^-$ | 0,*b*,*c*: 0,*b*,-*c* | | | | | ✓ | | ✓ | |
| | | | | 0,*b*,*c*: 0,-*b*,*c* | | ✓ | | | | | | ✓ |
| | | | | *a*,0,*c*: *a*,0,-*c* | | | | | | ✓ | ✓ | |
| | | | | *a*,0,*c*: -*a*,0,*c* | | ✓ | | | | | | ✓ |
| | | | | *a*,*b*,0: *a*,-*b*,0 | | ✓ | ✓ | | | | | |
| | | | | *a*,*b*,0: -*a*,*b*,0 | | ✓ | | ✓ | | | | |
| 3 | *I4/mmm* (#139) | 16 | $a^0b^+b^+$ | 0,*b*,*b*: 0,*b*,*b* | | ✓ | | | | | ✓ | |
| | | | | 0,*b*,-*b*:0,*b*,-*b* | | ✓ | | | | | ✓ | |
| | | | | *a*,0,*a*: *a*,0,*a* | | ✓ | | | | | ✓ | |
| | | | | *a*,0,-*a*: *a*,0,-*a* | | ✓ | | | | | ✓ | |
| | | | | *a*,*a*,0: *a*,*a*,0 | | ✓ | | | | | ✓ | |
| | | | | *a*,-*a*,0: *a*,-*a*,0 | | ✓ | | | | | ✓ | |
| *(5) | * *Immm* (#71) | 15 | $a^0b^+c^+$ | 0,*b*,*c*: 0,*b*,*c* | | ✓ | | | | | ✓ | |
| 8 | $R\bar{3}c$ (#167) | 14 | $a^-a^-a^-$ | *a*,*a*,*a*: -*a*,-*a*,-*a* | | | | | | | | |
| | | | | *a*,*a*,-*a*: -*a*,-*a*,*a* | | | | | | | | |
| | | | | *a*,-*a*,*a*: -*a*,*a*,-*a* | | | | | | | | |
| | | | | *a*,-*a*,-*a*: -*a*,*a*,*a* | | | | | | | | |
| 10 | *C2/c* (#15) | 13 | $a^-b^-b^-$ | *a*,*b*,*b*: -*a*,-*b*,-*b* | | | | | | | | |
| | | | | *a*,*b*,-*b*: -*a*,-*b*,*b* | | | | | | | | |
| | | | | *a*,*b*,*a*: -*a*,-*b*,-*a* | | | | | | | | |
| | | | | *a*,*b*,-*a*: -*a*,-*b*,*a* | | | | | | | | |
| | | | | *a*,*a*,*c*: -*a*,-*a*,-*c* | | | | | | | | |
| | | | | *a*,-*a*,*c*: -*a*,*a*,-*c* | | | | | | | | |
| 11 | $P\bar{1}$ (#2) | 12 | $a^-b^-c^-$ | *a*,*b*,*c*: -*a*,-*b*,-*c* | | | | | | | | |
| * (13) | * *Pnma* (#62) | 11 | $a^+a^-a^-$ | *a*,*a*,*a*: *a*,-*a*,-*a* | | | | | | | | |
| | | | | *a*,*a*,-*a*: *a*,-*a*,*a* | | | | | | | | |
| | | | | *a*,*a*,*a*: -*a*,*a*,-*a* | | | | | | | | |
| | | | | *a*,*a*,-*a*: -*a*,*a*,*a* | | | | | | | | |
| | | | | *a*,*a*,*a*: -*a*,-*a*,*a* | | ✓ | | | | | | ✓ |
| | | | | *a*,-*a*,*a*: -*a*,*a*,*a* | | ✓ | | | | | | ✓ |
| 13 | *Pnma* (#62) | 10 | $a^+b^-b^-$ | *a,b,b: a,-b,-b* | | | | | | | | |
| | | | | *a,b,-b: a,-b,b* | | | | | | | | |
| | | | | *a,b,a: -a,b,-a* | | | | | | | | |
| | | | | *a,b,-a: -a,b,a* | | | | | | | | |
| | | | | *a,a,c: -a,-a,c* | | ✓ | | | | | | ✓ |
| | | | | *a,-a,c: -a,a,c* | | ✓ | | | | | | ✓ |
| * (14) | * $P2_1/m$ (#11) | 9 | $a^+a^-c^-$ | *a,a,c: a,-a,-c* | | | | | | | | |
| | | | | *a,b,b: -a,b,-b* | | | | | | | | |
| | | | | *a,b,a: -a,-b,a* | | ✓ | | | | | | ✓ |
| 14 | $P2_1/m$ (#11) | 8 | $a^+b^-c^-$ | *a,b,c: a,-b,-c* | | | | | | | | |
| | | | | *a,b,c: -a,b,-c* | | | | | | | | |
| | | | | *a,b,c: -a,-b,c* | | ✓ | | | | | | ✓ |
| * (15) | * $P4_2/nmc$ (#137) | 7 | $a^+a^+a^-$ | *a,a,a: a,a,-a* | | | | | | | ✓ | |
| | | | | *a,-a,a: a,-a,-a* | | | | | | | ✓ | |
| | | | | *a,a,a: a,-a,a* | | ✓ | | | | | | |
| | | | | *a,a,-a: a,-a,-a* | | ✓ | | | | | | |
| | | | | *a,a,a: -a,a,a* | | ✓ | | | | | | |
| | | | | *a,a,-a: -a,a,-a* | | ✓ | | | | | | |
| * | * *Pmmn* (#59) | 6 | $a^+b^+b^-$ | *a,b,b: a,b,-b* | | | | | | | ✓ | |
| | | | | *a,b,a: a,b,-a* | | | | | | | ✓ | |
| | | | | *a,a,c: a,-a,c* | | ✓ | | | | | | |
| 15 | $P4_2/nmc$ (#137) | 5 | $a^+a^+c^-$ | *a,a,c: a,a,-c* | | | | | | | ✓ | |
| | | | | *a,-a,c: a,-a,-c* | | | | | | | ✓ | |
| | | | | *a,b,a: a,-b,a* | | ✓ | | | | | | |
| | | | | *a,b,-a: a,-b,-a* | | ✓ | | | | | | |
| | | | | *a,b,b: -a,b,b* | | ✓ | | | | | | |
| | | | | *a,b,-b: -a,b,-b* | | ✓ | | | | | | |
| * | * *Pmmn* (#59) | 4 | $a^+b^+c^-$ | *a,b,c: a,b,-c* | | | | | | | ✓ | |
| | | | | *a,b,c: a,-b,c* | | ✓ | | | | | | |
| | | | | *a,b,c: -a,b,c* | | ✓ | | | | | | |
| 4 | $Im\bar{3}$ (#204) | 3 | $a^+a^+a^+$ | *a,a,a: a,a,a* | | ✓ | | | | | ✓ | |
| | | | | *a,a,-a: a,a,-a* | | ✓ | | | | | ✓ | |
| | | | | *a,-a,a: a,-a,a* | | ✓ | | | | | ✓ | |
| | | | | *a,-a,-a: a,-a,-a* | | ✓ | | | | | ✓ | |
| * | * *Immm* (#71) | 2 | $a^+b^+b^+$ | *a,b,b: a,b,b* | | ✓ | | | | | ✓ | |
| | | | | *a,b,a: a,b,a* | | ✓ | | | | | ✓ | |
| | | | | *a,a,c: a,a,c* | | ✓ | | | | | ✓ | |
| 5 | *Immm* (#71) | 1 | $a^+b^+c^+$ | *a,b,c: a,b,c* | | ✓ | | | | | ✓ | |

**Table 4** Allowable planar defects (domain walls) in the $a^-a^-a^-$ tilt system. This table gives their characteristic symmetry operators W and the set of compatible tilt systems Λ(W). The octahedral tilt vector $\mathbf{t}_\mu$ describes the tilt of an octahedron in the $\mu$ crystal equivalent to one in the $\lambda$ crystal with tilt vector $\mathbf{t}_\lambda = [a, a, a]$, indicating which tilts must reverse across the interface. The vector $\mathbf{t}_{mean}$ gives the average of the two tilt systems at the interface.

| Defect | W | Λ(W) | $\mathbf{t}_\mu$ | $\mathbf{t}_{mean}$ | Local tilt system |
|---|---|---|---|---|---|
| APB (general) | ($\mathbf{I}$\|<100> + $\mathbf{u}$) | $a^0a^0a^0$ | $[-a, -a, -a]$ | $[0, 0, 0]$ | $a^0a^0a^0$ $Pm\bar{3}m$ |
| Twin (011) | ($m_{yz}$\|[000] + $\mathbf{u}$) | $a^0a^0a^0$<br>$a^0b^-b^-$ | $[-a, a, a]$ | $[0, a, a]$ | $a^0b^-b^-$ $Imma$ |
| Twin (001) | ($m_z$\|[000] + $\mathbf{u}$) | $a^0a^0a^0$<br>$a^0a^0c^-$<br>$a^-a^-a^0$ | $[-a, -a, a]$ | $[0, 0, a]$ | $a^0a^0c^-$ $I4/mcm$ |
| Twin (011) + APB | ($m_{yz}$\|[100] + $\mathbf{u}$) | $a^0a^0a^0$<br>$a^-b^0b^0$<br>$a^0b^-b^-$<br>$a^-a^-a^-$<br>$a^-b^-b^-$ | $[a, -a, -a]$ | $[a, 0, 0]$ | $a^-a^0a^0$ $I4/mcm$ |
| Twin (001) + APB | ($m_z$\|[010] + $\mathbf{u}$) | $a^0a^0a^0$<br>$a^-a^0a^0$<br>$a^-a^-a^0$ | $[a, a, -a]$ | $[a, a, 0]$ | $a^-a^-a^0$ $Imma$ |

**Table 5** Special planar defects on {001} planes displaced from the origin in the $a^-a^-a^-$ tilt system. This is similar to Table 4 but with the larger set of compatible tilt systems Γ(W).

| Defect | W | Γ(W) | $\mathbf{t}_\mu$ | Local tilt system |
|---|---|---|---|---|
| APB (001), displaced from origin | (**I**\|[001]) | $a^0a^0a^0$<br>$a^0a^0c^+$ | [-*a*, -*a*, -*a*] | $a^0a^0c^+$ *P4/mbm* |
| Twin (001) , displaced from origin | ($m_z$\|[001]) | $a^0a^0a^0$<br>$a^-a^0a^0$<br>$a^+a^0a^0$<br>$a^-a^-a^0$<br>$a^0b^-c^+$<br>$a^0b^+b^+$<br>$a^-a^-c^+$<br>$a^-b^-c^+$<br>$a^+a^-c^+$<br>$a^+b^-a^+$<br>$a^+b^-c^+$<br>$a^+a^+a^+$<br>$a^+b^+c^+$ | [*a*, *a*, -*a*] | $a^-a^-c^+$ *Pnma* |

## Acknowlegements

I would like to thank Prof. P.A. Thomas for her continued and enthusiastic support for this work, very useful discussions and the supply of $a^-a^-a^-$ $Na_{0.5}Bi_{0.5}TiO_3$, which was the motivation behind this analysis.